# Neutrino Mixing Discriminates Geo-reactor Models

S.T. Dye[1,2]

[1]*Department of Physics and Astronomy, University of Hawaii at Manoa, Honolulu, Hawaii, 96822 USA*
[2]*College of Natural Sciences, Hawaii Pacific University, Kaneohe, Hawaii, 96744 USA*
(Dated: 15 May 2009)

Geo-reactor models suggest the existence of natural nuclear reactors at different deep-earth locations with loosely defined output power. Reactor fission products undergo beta decay with the emission of electron antineutrinos, which routinely escape the earth. Neutrino mixing distorts the energy spectrum of the electron antineutrinos. Characteristics of the distorted spectrum observed at the earth's surface could specify the location of a geo-reactor, discriminating the models and facilitating more precise power measurement. The existence of a geo-reactor with known position could enable a precision measurement of the neutrino oscillation parameter $\Delta m_{21}^2$.

PACS numbers: 14.60.Pq, 91.35.-x, 93.85.Pq, 28.50.Ft

The accumulation of actinide elements in the interior of the earth can lead to sustained nuclear fission reactions [1]. Fission products undergo beta decay with the emission of electron antineutrinos, which routinely escape the earth. Isotopic analysis of uranium deposits confirms that nuclear reactors occurred naturally near the surface of the earth ~1.7 Ga ago [2]. Hypotheses for presently existing natural breeder reactors propose deep-earth locations, including the center of the core [3], the interface between the inner and outer core [4], and the core-mantle boundary [5]. These geo-reactor models suggest reactor output power sufficient to explain terrestrial heat flow measurements [6] and helium isotope ratios in oceanic basalts [7]. The measurement of electron antineutrinos from commercial nuclear reactors leads to an experimental upper limit to the power of an earth-centered geo-reactor of 6.2 TW (90% C.L.) [8]. Allowing for possible locations along a diameter through the core leads to a power limit spanning 1.3-15 TW. Uncertainty in the location of the geo-reactor leads to uncertainty in the power estimate.

This report describes a method for locating deep-earth geo-reactors, if they exist, by measuring distortions of the electron antineutrino energy spectrum at the surface of the earth. These spectral distortions result from the mixing of neutrino mass states along the path from source to detector [9]. The distortion pattern specifies the length of the path, thereby offering the potential to discriminate geo-reactor models.

An empirical fit estimates the energy spectrum of electron antineutrinos detected near a reactor with the formula $N(E_{\bar{\nu}_e}) \propto (E_{\bar{\nu}_e} - 1.4)^2 \exp\!-\!\left(\dfrac{E_{\bar{\nu}_e} + 0.8}{3.2}\right)^2$, valid for $E_{\bar{\nu}_e}$ greater than the interaction threshold energy, with all quantities in MeV. The decaying exponential describes antineutrino production and the rising quadratic represents the interaction cross section. Determining the detection rate from an earth-centered geo-reactor follows from previous work, which designates 8.08 (TW·y·$10^{32}$ free-protons)$^{-1}$ antineutrinos with energy greater than 3.4 MeV detected with an efficiency of 1.0 [10].

Mixing of neutrino mass states along the flight path converts some of the electron antineutrinos to a flavor that does not participate in the detection mechanism. With mixing between only two mass states, the probability that an electron antineutrino does not convert is $P_{\bar{\nu}_e \to \bar{\nu}_e} \cong 1 - \sin^2(2\theta_{12})\sin^2(1.27\Delta m_{21}^2 L/E_{\bar{\nu}_e})$ with $L$ the length of the neutrino flight path in meters, $E_{\bar{\nu}_e}$ the energy of the neutrino in MeV, $\Delta m_{21}^2$ the difference of the squares of the neutrino mass states, and $\theta_{12}$ the mixing angle between mass states [9]. This study employs recently measured values of the neutrino mixing parameters $\Delta m_{21}^2$ and $\theta_{12}$ derived from the detection of antineutrinos from commercial power reactors [8].

The present method of detecting reactor antineutrinos employs essentially the same technology used to first observe their interactions more than five decades ago [11]. It takes advantage of the signal coincidence and relatively large interaction cross section of inverse neutron beta decay $\bar{\nu}_e + p \to e^+ + n$. Each product initiates a detectable signal in scintillating liquid, forming a spatial and temporal coincidence. Scintillation light intercepted by inward-looking photomultiplier tubes estimates the electron antineutrino energy from the positron signal and verifies the interaction from the subsequent neutron signal. This detection method provides accurate estimates of the energies of reactor antineutrinos interacting with free protons [8].

The top panel of Figure 1 shows the unmixed and mixed energy spectra of electron antineutrinos from an earth-centered geo-reactor. These idealized spectra contain the number of events detected from an exposure of TW·y·$10^{33}$ free-protons. Note the increasing spacing of the spectral distortions with increasing energy. The lower panel of Figure 1 shows the spectra resulting from the same exposure to a geo-reactor located at the near point of the core-mantle boundary. Note that the spectral distortions due to neutrino mixing have greater spacing for the closer geo-reactor than those for the farther geo-reactor. These idealized spectra do not take into account statistical fluctuations inherent in a spectrum of detected events.

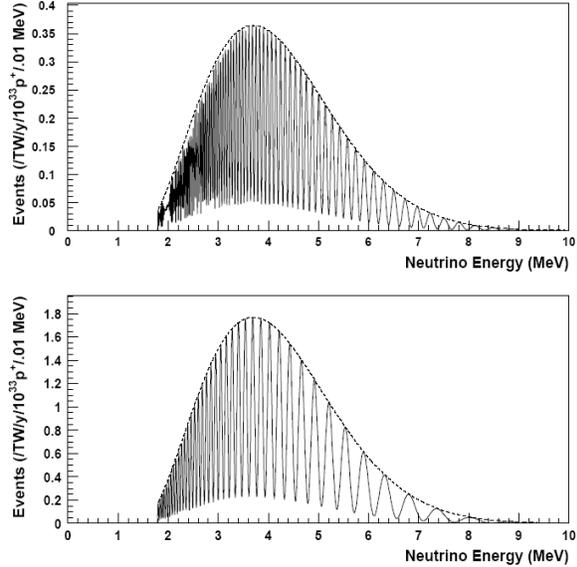

FIG. 1: The upper (lower) panel shows the energy spectra that would be detected from a geo-reactor located at a distance of 6371 km (2891 km) from the detector. Dashed curve at the top is the unmixed spectrum. Solid curve underneath is modulated by neutrino mixing. All spectra begin at 1.8 MeV, which is the threshold energy for inverse neutron beta decay.

Assessing the viability of locating a deep-earth geo-reactor by measuring the spectral distortions of electron antineutrinos requires consideration of detector energy resolution. A currently operating detector of reactor antineutrinos achieves a fractional uncertainty in measured energy of $\delta E/E = 6.5\%$ at one standard deviation [8]. Improvements to the energy resolution of this detector would result from increases to the fractional area of photocathode [12], the photocathode quantum efficiency [13], and the light output of the scintillating liquid [14]. Realizing the increases in sensitivity afforded by present technology would improve the energy resolution of such a detector by at least a factor of two.

The left (right) panels of Figure 2 show the idealized energy spectra of electron antineutrinos detected with a resolution of 6% (3%) due to a solitary geo-reactor located at the center of the earth, the near point of the interface between the inner and outer core, and the near point of the core-mantle boundary. These spectra obtain from distributing the contents of each energy bin to a Gaussian with standard deviation equal to the energy resolution. Note the greater amplitude of distortions in the energy spectra detected with 3% resolution. The measurement of mixing-induced distortions in the spectra of detected reactor antineutrinos benefits from improved energy resolution at deep-earth distances.

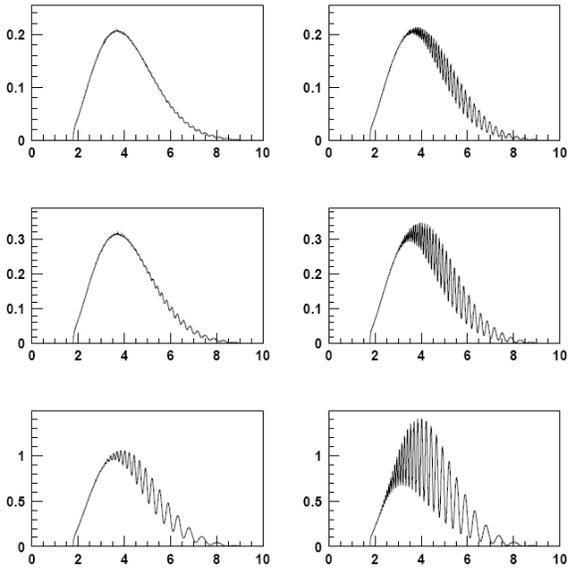

FIG. 2: The left (right) panels show electron antineutrino energy spectra detected with energy resolution of 6% (3%) from a solitary geo-reactor located at a distance of 6371 km (top), 5149 km (middle), and 2891 km (bottom). Plot axes are the same as Figure 1.

An assessment of the ability to determine the electron antineutrino source distance by measuring spectral distortions utilizes the Rayleigh test [15]. This statistical test returns the power $P$ of spectral distortions for an assumed length of neutrino flight path. The probability that the spectral distortions are due to a random fluctuation is $e^{-P}$ times the number of independent distances tested. The width of the reactor spectrum (1.8-9.0 MeV) determines the independent distance $\frac{2\pi}{1.27\Delta m_{21}^2}\frac{E_{max}E_{min}}{E_{max}-E_{min}}$, which is 147 km in this study. A search for a source calculates the power over a range of distances from the detector. A peak in the power distribution indicates increased likelihood that the spectral distortions are due to a source at the corresponding distance. The statistical test assumes the measured value of the mixing parameter $\Delta m_{21}^2$. Assuming a larger value underestimates the source distance and vice versa. The fractional uncertainty in the measured value of $\Delta m_{21}^2$ is ±2.8% (68% CL) [8]. This limits the precision of the distance determination.

Figure 3 shows the power distributions for the idealized spectra displayed in Figure 2. The panels on the left (right) correspond to detector energy resolution of 6% (3%). Note that the power distributions have peaks within about 20 km of the correct source distances. These peaks are more pronounced in the spectra measured with 3% resolution than those measured with 6% resolution. Since the energy spectra do not take in to account statistical sampling of recorded events, the power distributions do not represent actual measurements and serve only to illustrate the potential of this method.

The foregoing discussion concentrates on locating a solitary fission reactor. A geo-reactor could consist of a number of individual fission sites. The top panel of Figure 4 shows the idealized energy spectrum detected with 3% resolution resulting from four deep-earth reactors of equal power at different locations along a diameter of the core. These four positions correspond to the near point to the core-mantle boundary, the near and far points to the interface between the inner and outer core, and the center of the earth. The bottom panel of Figure 4 shows the power distribution for this spectrum. Note the peaks in the distribution are at distances matching the source positions. The width of the peaks, which is about 1000 km, determines the resolution. This method is potentially capable of resolving multiple, simultaneous sources of reactor antineutrinos, provided the detector-source distances have separations of ~500 km or more.

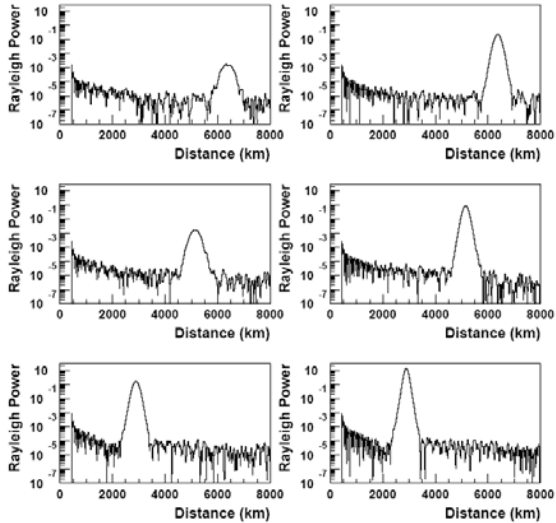

FIG 3: Rayleigh power distributions as a function of distance for the spectra in Figure 2. Left (right) panels are for the spectra with 6% (3%) energy resolution. Rising power at smaller distances is due to the shape of the reactor spectrum.

A more stringent test of this method applies to continuous distributions of fission sites. The left panels of Figure 5 show the idealized energy spectra detected with 3% resolution resulting from deep-earth reactors distributed uniformly on geocentric, spherical shells of different radii. The right panels of Figure 5 show the power distributions for these spectra. Coherence of the spectral distortions persists out to a radius of at least 50 km, which is sufficient for testing the earth-centered geo-reactor model [3].

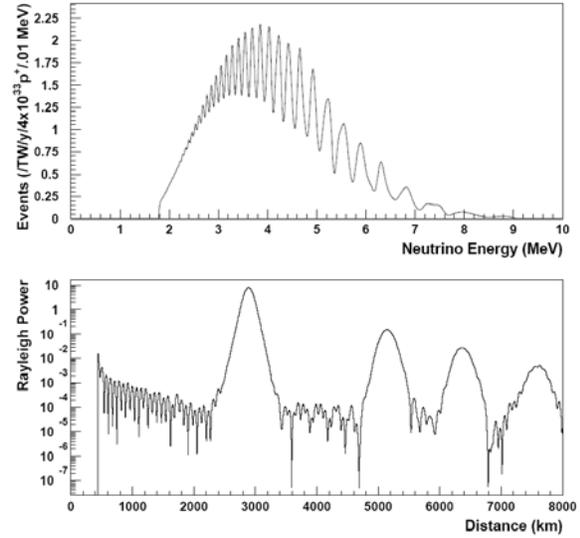

FIG. 4: The top panel shows the energy spectrum detected with 3% resolution resulting from sources of equal power at distances of 2891 km, 5149 km, 6371 km, and 7593 km. The bottom panel shows the power distribution for this spectrum.

Assessing the exposure required to locate the distance to a source of reactor antineutrinos involves random sampling of the idealized spectra. The sample total represents the number of event detections resulting from a given exposure. Calculating the Rayleigh power distribution of an event spectrum yields the most likely detector-source distance. A successful trial locates the source within one independent distance. Repeating many trials evaluates the efficiency for locating a source distance for a given exposure.

Figure 6 shows the efficiencies for locating a source at three different distances as a function of exposure. The source position for the upper points is the center of the earth (6371 km), for the middle points is the near point to the interface between the inner and outer core (5149 km), and for the lower points is the core-mantle boundary (2981 km). For an efficiency of 95%, these distances require exposures of 20, 6, and 0.4 TW-y-$10^{33}$p$^+$, respectively. With 1.5 times these exposures the efficiency climbs to 99% and the standard deviation of the distance distributions becomes comparable to the independent distance. Exposures of this magnitude with 3% energy resolution are within technological capability.

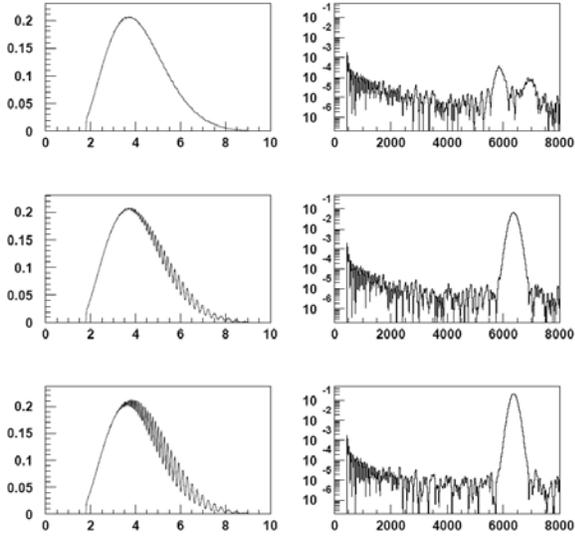

FIG. 5: The left panels show the energy spectra detected with 3% resolution resulting from geo-reactors distributed uniformly on an earth-centered spherical shell of radius 500 km (top), 50 km (middle), and 5 km (bottom). Plot axes are the same as Figure 1. The right panels show the corresponding power distributions for these spectra. Plot axes are the same as Figure 3.

This study does not incorporate potential sources of background, which include commercial nuclear reactors, cosmic radiation, and terrestrial antineutrinos [16]. Situating the detector at a location with an overburden of several thousand m.w.e. that is far from continents would minimize background from these sources. A deep-ocean antineutrino observatory could provide these attributes [17].

Measuring distortions in the energy spectrum of electron antineutrinos induced by mixing of mass states is a viable method for specifying the path length of neutrinos. This method could estimate the distance to a deep-earth geo-reactor with an uncertainty comparable to the measurement error of the neutrino mixing parameter $\Delta m_{21}^2$. Knowing the distance to the geo-reactor would afford a precise estimate of the output power. Distance estimates from several earth-surface locations could determine the location of a solitary geo-reactor, thereby discriminating geo-reactor models. This method is capable of estimating the distances to multiple sources, which could prove useful to nuclear nonproliferation. Indeed, a practical application of this method may involve locating nuclear reactors at the surface of the earth. Detector energy resolution is crucial when applying this method to source distances comparable to the radius of the earth. If a deep-earth geo-reactor were to exist at a well-defined location, such as the center of the earth, this method could contribute to a more precise measurement of the neutrino mixing parameter $\Delta m_{21}^2$.

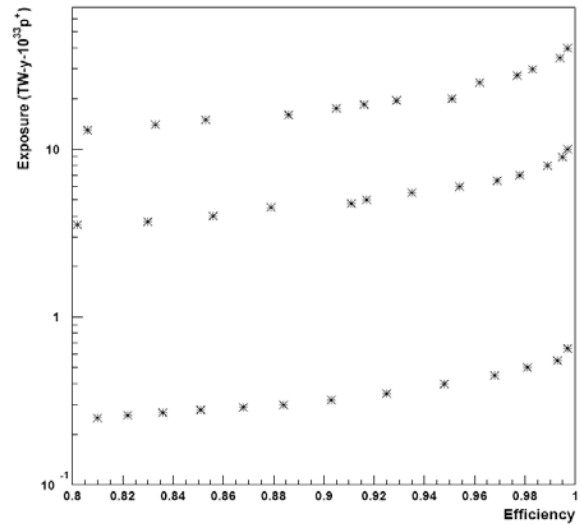

FIG. 6: Efficiencies for locating source distances as a function of exposure. The upper, middle, and lower points are for a source at 6371 km, 5149 km, and 2981 km, respectively.

The author thanks J.G. Learned for discussions, S. Matsuno for help with PAW,

and J.M. Mahoney for comments and the helium isotope reference. This work was supported in part by the Hawaii Pacific University Trustees' Scholarly Endeavors Program.